\documentclass[]{revtex4}
\usepackage{graphicx}

\begin{document}

\title{Neutrino astrophysics: A new tool for exploring the universe\footnote{Perspective article, Science Special Issue on "particle astrophysics" (Jan 2007)}}

\author{Eli Waxman\\Physics faculty, Weizmann institute of science, Rehovot 76100, Israel}

\maketitle

In the past four decades a new type of astronomy has emerged, where instead of looking up into the sky "telescopes" are buried miles underground or deep under water or ice and search not for photons (that is, light), but rather for particles called neutrinos. Neutrinos are nearly massless particles that interact very weakly with matter. The detection of neutrinos emitted by the Sun and by a nearby supernova provided direct tests of the theory of stellar evolution and led to modifications of the standard model describing the properties of elementary particles. At present, several very large neutrino detectors are being constructed, aiming at the detection of the most powerful sources of energy and particles in the universe. The hope is that the detection of neutrinos from these sources, which are extra-Galactic and are most likely powered by mass accretion onto black-holes, will not only allow study of the sources, but, much like solar neutrinos, will also provide new information about fundamental properties of matter.

\bigskip

Neutrino astronomy was initiated as an attempt to provide a direct experimental test for the theory of stellar evolution \citep{jnb_book}. According to this theory the Sun is powered by the nuclear fusion of hydrogen into helium, which takes place deep in the solar interior. The mass of four H atoms is larger than that of the He atom into which they fuse. The excess mass $m$ is converted to energy, according to $E=mc^2$ ($c$ is the speed of light), which keeps the Sun shining. It was suggested in the mid 1960s that one could test this model by searching for neutrinos, which were predicted to be emitted by the fusion process. Unlike photons that are emitted from the Sun's surface, the weak interaction of neutrinos with matter allows them to escape from the Sun's core and directly reach detectors on Earth. 

The weak interaction of neutrinos with matter also implies that they are very difficult to detect, requiring the construction of detectors with several kilo-tons of detecting medium. Although the probability that a neutrino passing through kilo-tons of matter would interact within the detector, or be "captured," is very small, the large flux of neutrinos from the Sun, some 100 billion neutrinos per square centimeter per second, allows hundreds of them to be detected every year. In addition to being very massive, all detectors are also buried deep underground. At the surface of Earth there is a large flux of high energy particles. Such particles are produced mainly by the interaction of cosmic-rays, high energy particles produced in space, with the atmosphere. Penetration of high energy particles into the detector may lead to interactions that would mimic neutrino interactions. Burial of the detector deep underground suppresses this background, because only neutrinos can penetrate deep enough into Earth to reach the detector. 

The detection of solar neutrinos was an impressive confirmation of the hypothesis of a nuclear fusion origin of stellar energy. However, it also posed a challenge: The measured neutrino flux was roughly 1/2 that predicted by theory. Shortly after this discrepancy was first reported in 1968, it was suggested to be due to shortcomings of the standard model describing the properties of elementary particles \citep{Bilenky}. Neutrinos come in three types, or "flavors": electron-type ($\nu_e$), muon-type ($\nu_\mu$(, and tau-type ($\nu_\tau$). It was proposed that $\sim1/2$ of the neutrinos which are produced in the Sun by nuclear fusion, and are all of electron-type, change their flavor to $\nu_\mu$ or $\nu_\tau$ as they propagate to Earth. Such flavor conversion, commonly termed "oscillation," was not expected according to the standard model and would explain why neutrino detectors sensitive to $\nu_e$ only would miss $\sim1/2$ the solar neutrino flux.

The oscillation explanation was confirmed in 2001 with the detection of  the "missing" $\nu_e$ flux in the form of $\nu_\mu$ and $\nu_\tau$ flux by an experiment sensitive to all flavors \citep{GGN}. 
Independent evidence for neutrino oscillations came from measurements of atmospheric neutrinos, produced by cosmic-ray interactions in the atmosphere, which indicate conversion of $\nu_\mu$ to $\nu_\tau$ \citep{GGN}. Neutrino oscillations are the first, and so far only, experimental phenomenon not accounted for by the standard model. It is most naturally explained by a model in which 3 neutrinos with different masses exist-- say $\nu_1$, $\nu_2$ and $\nu_3$ with masses $m_1$, $m_2$ and $m_3$ respectively-- and in which neutrinos of different flavors are in fact composed of different mixtures of $\nu_1$, $\nu_2$ and $\nu_3$. $\nu_e$, for example, is a roughly equal mixture of $\nu_1$ and $\nu_2$ with little (if any) contribution of $\nu_3$. 

After the discovery of neutrino oscillations by observing natural (solar and atmospheric) neutrino sources, oscillations were also measured with neutrinos produced in nuclear reactors and particle accelerators. Oscillation measurements provide constraints on the neutrino "mixing parameters" \citep{Bilenky,GGN}, that is on the composition (in terms of $\nu_{1,2,3}$) of neutrinos of different flavors, and on the mass-squared differences, $m^2_{2}-m_1^2=8\times10^{-5}({\rm eV}/c^2)^2$ and $|m_3^2-m^2_{2}|=2\times10^{-3}({\rm eV}/c^2)^2$. Here, masses are given in energy units, where $m=E/c^2$; 1~eV is the typical binding energy of molecules and corresponds to roughly one millionth of the electron mass, $m_e c^2=0.5\times10^6$~eV. Oscillation experiments can not determine the absolute values of the masses, and current data do not allow one to discriminate between the two "hierarchies," $m_1<m_2<m_3$ and $m_3<m_1<m_2$. An upper limit on the mass of the most massive neutrino, $m\le2~{\rm eV}/c^2$, is set by measurements of radioactive decay of Tritium \citep{nu_mass}. A similar upper limit is obtained from surveys of the large scale distribution of galaxies: The universe is filled with a "neutrino background," a relic of the big bang, and if neutrinos were too massive they would have suppressed the formation of large scale structures in the universe \citep{nu_cosmo}. 

A model explaining the origin of neutrino masses and mixing does not yet exit \citep{GGN}. In order to make progress towards such a model, large radioactive decay experiments are planned in order to measure the absolute neutrino mass scale \citep{nu_mass}, and large oscillation experiments involving reactors and specially designed accelerator beams are planned for determining the mass hierarchy (and for accurate determination of the mixing parameters) \citep{ter_nu_exp}. These experiments will also try to ascertain whether the mixing properties of neutrinos and of their anti particles, anti neutrinos, are identical. Answering the above questions would be important not only for the construction of a model accounting for neutrino mass and mixing. It may also be relevant for answering another open question-- why our universe appears to be composed mainly of particles and not of anti particles \citep{leptogen}. 

According to the theory of stellar evolution, stars more massive than the Sun by a factor of 10 or more end their lives with an explosion, a supernova, that ejects most of the star's mass and leaves behind a dense "neutron star" remnant of roughly one solar mass. Theory predicts that most of the energy generated by the explosion would be carried away from the star by neutrinos. This prediction was confirmed \citep{jnb_book} with the detection in 1987 of neutrinos emitted by the supernova SN~1987A, which exploded in the Large Magellanic Cloud, a small satellite galaxy of our own Galaxy, the Milky Way, lying at a distance of some 150 thousand light years. 

The characteristic energy of neutrinos produced in the Sun or in supernova explosions is on the order of megaelectron volts (1~MeV$=10^6$eV), which is the characteristic energy released in the fusion or fission of atomic nuclei. The detection of MeV neutrinos from sources well outside our local Galactic neighborhood, at distances ranging from several million light years (the typical distance between galaxies) to several billion light years (the size of the observable universe) is impossible with present techniques. To extend the distance accessible to neutrino astronomy to the edge of the observable universe, several high energy neutrino telescopes are currently being constructed deep under ice or water. These telescopes are designed for the detection of neutrinos with energies exceeding terraelectron volts (1~TeV$=10^{12}$~eV) and are planned to reach effective masses exceeding  1~gigaton \citep{Halzen}. 
\begin{figure}[htbp]
\includegraphics[width=10cm]{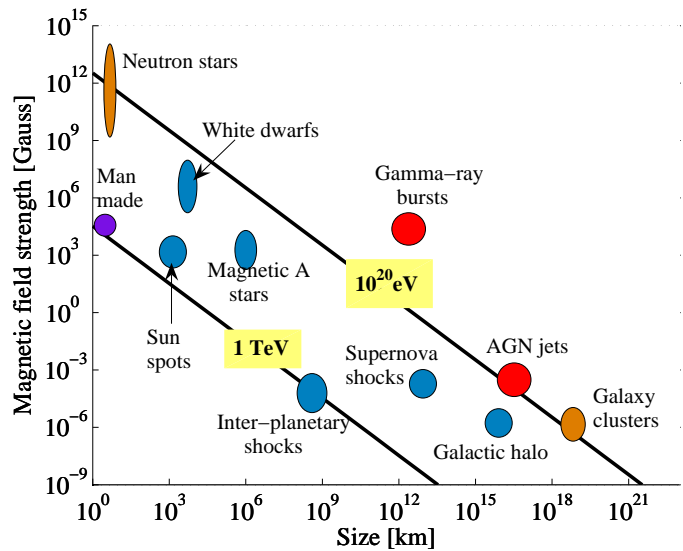}
\caption{Charged particles are confined to their astrophysical accelerators by magnetic fields. Magnetic confinement requires the product of field strength and accelerator size to exceed a value, which increases with particle energy. The figure shows the size and magnetic field strength of possible sites of particle acceleration. 
(The magnetic field is measured in Gauss units, where the Earth's magnetic field is $\sim1$~G.) Proton acceleration to 1~TeV or $10^{20}$~eV is possible only for sources lying above the appropriately marked lines. This is a necessary, but not sufficient requirement: Proton acceleration to $10^{20}$~eV is impossible in galaxy clusters (because the acceleration time in these objects is larger than the age of the universe) and unlikely in highly magnetized neutron stars (due to severe energy losses). The characteristics of terrestrial man-made accelerators, which are planned to reach $\sim1$~TeV, are shown for comparison.}
\label{fig:hillas}
\end{figure}
\begin{figure}[htbp]
\includegraphics[width=8cm]{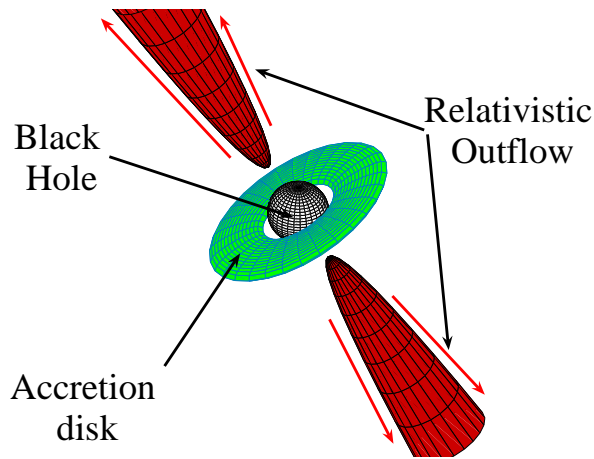}
\caption{GRBs and AGN are believed to be powered by black holes. The accretion of mass onto the black hole, through an accretion disk, releases large amounts of gravitational energy. If the black hole is rotating rapidly, the rotational energy may also be released by slowing the black hole through interaction with the disk. The energy released drives a jet-like relativistic outflow. The observed radiation is produced as part of the energy carried by the jets is converted, at large distance from the central black hole, to electromagnetic radiation. GRBs are believed to be powered by $\sim1$ solar mass black holes with jets extending to distances larger than the size of the solar system, producing short (typically 1 to 100~s long) flashes of luminosity exceeding that of the Sun by 19 orders of magnitude. AGN are powered by million to billion Solar mass black holes residing at the centers of distant galaxies, with jets extending to distances larger than the size of a galaxy, producing a steady luminosity exceeding that of the Sun by 12 orders of magnitude.}
\label{fig:BHjet}
\end{figure}

The sources targeted by high energy ($\ge1$~TeV) neutrino detectors are "cosmic accelerators," in which particles are accelerated to extreme energies. The existence of cosmic-rays, high energy particles that are produced in astrophysical objects and are observed as they hit and interact with the Earth's atmosphere, has been mentioned above. The sources of these particles have not yet been identified, and the mechanisms that lead to particle acceleration are not well understood. One of the major goals of $\ge1$~TeV neutrino detectors is to resolve these open questions.

Particle acceleration theories are most challenged by the highest energy particles observed \citep{W_nobel04}. These particles are most likely protons and their energy exceeds $10^{20}$~eV, or 100 million TeV. Althuogh there are a variety of astrophysical objects suspected of being "cosmic accelerators" (fig.~\ref{fig:hillas}), only two types of sources are known that may be capable of accelerating protons to $10^{20}$~eV: gamma-ray bursts (GRBs) and active galactic nuclei (AGN). These objects lie at cosmological distances, billions of light years away, and are the brightest known objects in the universe (fig.~\ref{fig:BHjet}). Although GRB and AGN models are generally successful in explaining most observations, they are largely phenomenological and major questions remain open. These include the mechanisms by which gravitational energy is harnessed to power the sources, and the mechanism of particle acceleration.

A direct association of cosmic-rays with their sources is difficult: Magnetic fields in interstellar and intergalactic space deflect the electrically charged cosmic-rays, which, therefore, do not travel on straight lines and do not point back to their sources. Neutrinos, on the other hand, are electrically neutral and therefore travel at straight lines and do point back to their sources. Whatever the cosmic accelerators are, they are expected to be sources of high energy neutrinos and therefore to be identifiable by their neutrino emission. This expectation is based on the fact that the interaction of high energy cosmic-rays with radiation or matter leads to the production of neutrinos. High energy protons, for example, may interact with photons to produce pions, particles that decay and produce muon and electron neutrinos.

\begin{figure}[htbp]
\includegraphics[width=10cm]{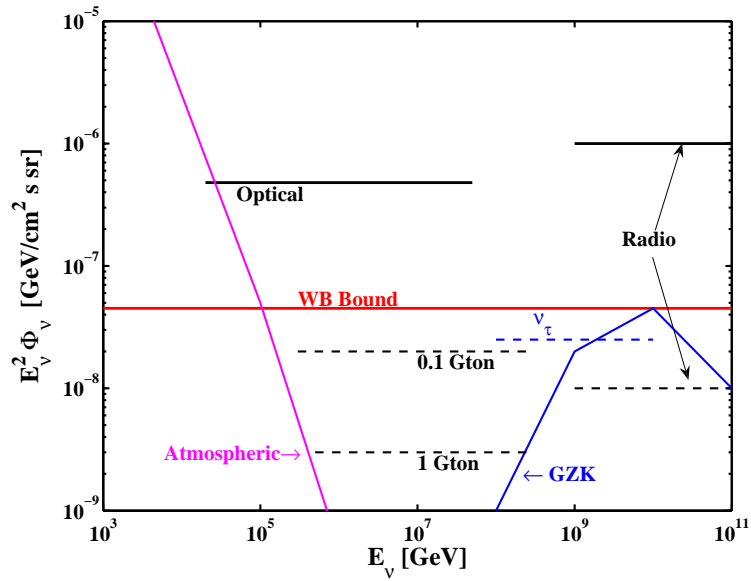}
\caption{The cosmic-ray upper bound on the extra-Galactic high energy neutrino intensity ($\nu_\mu+\nu_\tau$ assuming $\nu_e:\nu_\mu:\nu_\tau=1:1:1$ flux ratios), compared with experimental upper bounds (solid lines) provided by optical detectors under water or ice and by radio detectors, and with the expected sensitivity (dashed lines) of various detectors: 0.1~gigaton and 1~gigaton under water/ice optical detectors, radio detectors, and ground arrays of particle detectors (sensitivity to $\nu_\tau$). The intensity $\Phi_\nu$ is the number of neutrinos of given energy $E_\nu$ (measured in GeV=1000 MeV) crossing in 1 second a unit area ($1{\rm cm^2}$) of a detector observing a solid angle of 1~steradian of the sky. A detailed description of the experiments is given in~\citep{Halzen}. The curve marked GZK shows the neutrino intensity expected to be produced by the interaction of high energy cosmic-ray protons with the cosmic microwave background, the relic radiation of the big bang. Also shown is the atmospheric neutrino intensity, which is produced by cosmic ray interactions in the atmosphere and constitutes the main background.}
\label{fig:WBbound}
\end{figure}
Observations of high energy cosmic-rays provide a means to estimating the expected high energy neutrino flux and hence the detector size required to measure it. The observed cosmic-ray flux sets an upper bound to the neutrino flux produced by extra-Galactic sources \citep{W_nobel04}, which implies that gigaton neutrino telescopes are needed to detect the expected extra-Galactic flux in the energy range of $\sim1$ to $\sim1000$~TeV, and much larger effective mass is required at higher energy (fig.~\ref{fig:WBbound}). A flux comparable to the bound at $\sim1$ to $\sim1000$~TeV would produce hundreds of events per year in a gigaton detector. A few tens of events per year are expected in a gigaton telescope if GRBs are the sources of high energy protons. These events will be correlated in time and direction with GRB photons, allowing for an essentially background free experiment.

Detection of high energy neutrinos with the next generation of telescopes will probe the most powerful cosmic accelerators, including GRBs and AGN, and will allow study of the physical mechanisms powering them. It will also provide new tests of neutrino oscillation theory and probes of fundamental physics that are not available with terrestrial man made sources: Flavor measurements of high energy neutrinos will contribute to the determination of the mixing parameters (e.g. to resolving the mass hierarchy ambiguity and to testing for differences in particle and anti-particle behavior) \citep{Winter06}. The angular dependence of neutrino detection rate may allow testing for deviations from standard model predictions of the neutrino-nucleon interaction cross section at energies not accessible to terrestrial accelerators \citep{Halzen}. Detection of neutrinos from GRBs could be used to test the simultaneity of neutrino and photon arrival to an accuracy of $\sim1$~s. This would the validity of the underlying assumption of special relativity-- that photons and neutrinos have the same limiting speed-- to be determined with an accuracy of one part in $10^{17}$, and the validity of the weak equivalence principle-- the basic assumption of general relativity according to which photons and neutrinos should experience the same time delay as they pass through a gravitational potential-- to be measured with an accuracy better than one part in $10^6$ \citep{W_nobel04}. Previous applications of these ideas to supernova 1987A yielded much weaker upper limits, on the order of $10^{-8}$ and $10^{-2}$, respectively \citep{jnb_book}. Finally, neutrino telescopes may contribute to the detection of "dark matter," unseen particles that were not detected in laboratories on Earth and are believed to contain most of the mass in the universe \citep{Sadoulet}, through the detection of neutrinos produced by annihilation of dark matter particles.


\begin{thebibliography}{}

\bibitem[Bahcall(1989)]{jnb_book}
  J. N. Bahcall, {\it Neutrino Astrophysics} (Cambridge Univ. Press, NY 1989).
\bibitem[Bilenky(2005)]{Bilenky} 
  Bilenky, S.~M.\ 2005, Physica Scripta Volume T, 121, 17 (Proc. Nobel Symp. 129: Neutrino physics) 
\bibitem[Gonzalez-Garcia \& Nir(2003)]{GGN} 
  Gonzalez-Garcia, M.~C., \& Nir, Y.\ 2003, Reviews of Modern Physics, 75, 345
\bibitem[Weinheimer(2005)]{nu_mass} 
  Weinheimer, C.\ 2005, Physica Scripta Volume T, 121, 166 (Proc. Nobel Symp. 129: Neutrino physics)
\bibitem[Elgar{\o}y \& Lahav(2005)]{nu_cosmo} 
  Elgar{\o}y, {\O}., \& Lahav, O.\ 2005, New Journal of Physics, 7, 61 
\bibitem[Lindner(2005)]{ter_nu_exp} 
  Lindner, M.\ 2005, Physica Scripta Volume T, 121, 78 (Proc. Nobel Symp. 129: Neutrino physics) 
\bibitem[Yanagida(2005)]{leptogen} 
  Yanagida, T.\ 2005, Physica Scripta Volume T, 121, 137 (Proc. Nobel Symp. 129: Neutrino physics) 
\bibitem[Halzen(2006)]{Halzen} 
  Halzen, F.\ 2006, this volume
\bibitem[Waxman(2005)]{W_nobel04} 
  Waxman, E.\ 2005, Physica Scripta Volume T, 121, 147 (Proc. Nobel Symp. 129: Neutrino physics) 
\bibitem[Winter(2006)]{Winter06} 
  Winter, W.\ 2006, \prd, 74, 033015 
\bibitem[Sadoulet(2006)]{Sadoulet} 
  Sadoulet, B.\ 2006, this volume

\end{thebibliography}
\end{document}